\input harvmac
\def\journal#1&#2(#3){\unskip, \sl #1\ \bf #2 \rm(19#3) }
\def\andjournal#1&#2(#3){\sl #1~\bf #2 \rm (19#3) }

\def\ie{{\it i.e.}}
\def\eg{{\it e.g.}}

\def\frac#1#2{{#1\over#2}}

\def\half{\frac12}

\def\inbar{\,\vrule height1.5ex width.4pt depth0pt}
\def\IC{\relax\hbox{$\inbar\kern-.3em{\rm C}$}}
\def\IR{\relax{\rm I\kern-.18em R}}
\def\IP{\relax{\rm I\kern-.18em P}}
\def\IZ{\relax{\rm I\kern-.18em Z}}

%
%

%
\catcode`\@=11
\def\slash#1{\mathord{\mathpalette\c@ncel{#1}}}
\overfullrule=0pt

\def\SS{{\cal S}}

\def\underrel#1\over#2{\mathrel{\mathop{\kern\z@#1}\limits_{#2}}}

\catcode`\@=12


%

\def\det{{\rm det}}

\def \sinh{{\rm sinh}}
\def \cosh{{\rm cosh}}

\def\det{{\rm det}}
\def\exp{{\rm exp}}



\lref\SenTM{
A.~Sen,
``Dirac-Born-Infeld action on the tachyon kink and vortex,''
arXiv:hep-th/0303057.
}

\lref\SenQA{
A.~Sen,
``Time and tachyon,''
arXiv:hep-th/0209122.
}

\lref\SenAN{
A.~Sen,
``Field theory of tachyon matter,''
Mod.\ Phys.\ Lett.\ A {\bf 17}, 1797 (2002)
[arXiv:hep-th/0204143].
}

\lref\SenIN{
A.~Sen,
``Tachyon matter,''
JHEP {\bf 0207}, 065 (2002)
[arXiv:hep-th/0203265].
}

\lref\SenNU{
A.~Sen,
``Rolling tachyon,''
JHEP {\bf 0204}, 048 (2002)
[arXiv:hep-th/0203211].
}

\lref\LambertZR{
N.~Lambert, H.~Liu and J.~Maldacena,
``Closed strings from decaying D-branes,''
arXiv:hep-th/0303139.
}

\lref\LarsenWC{
F.~Larsen, A.~Naqvi and S.~Terashima,
``Rolling tachyons and decaying branes,''
JHEP {\bf 0302}, 039 (2003)
[arXiv:hep-th/0212248].
}

\lref\GarousiTR{
M.~R.~Garousi,
``Tachyon couplings on non-BPS D-branes and Dirac-Born-Infeld action,''
Nucl.\ Phys.\ B {\bf 584}, 284 (2000)
[arXiv:hep-th/0003122].
}

\lref\SenMD{
A.~Sen,
``Supersymmetric world-volume action for non-BPS D-branes,''
JHEP {\bf 9910}, 008 (1999)
[arXiv:hep-th/9909062].
}

\lref\LeblondDB{
F.~Leblond and A.~W.~Peet,
``SD-brane gravity fields and rolling tachyons,''
arXiv:hep-th/0303035.
}

\lref\BerkoozJE{
M.~Berkooz, B.~Craps, D.~Kutasov and G.~Rajesh,
``Comments on cosmological singularities in string theory,''
arXiv:hep-th/0212215.
}

\lref\HarveyWM{
J.~A.~Harvey, D.~Kutasov, E.~J.~Martinec and G.~Moore,
``Localized tachyons and RG flows,''
arXiv:hep-th/0111154.
}

\lref\KutasovAQ{
D.~Kutasov, M.~Marino and G.~W.~Moore,
``Remarks on tachyon condensation in superstring field theory,''
arXiv:hep-th/0010108.
}

\lref\KutasovQP{
D.~Kutasov, M.~Marino and G.~W.~Moore,
``Some exact results on tachyon condensation in string field theory,''
JHEP {\bf 0010}, 045 (2000)
[arXiv:hep-th/0009148].
}

\lref\HarveyNA{
J.~A.~Harvey, D.~Kutasov and E.~J.~Martinec,
``On the relevance of tachyons,''
arXiv:hep-th/0003101.
}

\lref\MarinoQC{
M.~Marino,
``On the BV formulation of boundary superstring field theory,''
JHEP {\bf 0106}, 059 (2001)
[arXiv:hep-th/0103089].
}

\lref\NiarchosSI{
V.~Niarchos and N.~Prezas,
``Boundary superstring field theory,''
Nucl.\ Phys.\ B {\bf 619}, 51 (2001)
[arXiv:hep-th/0103102].
}

\lref\WittenCR{
E.~Witten,
``Some computations in background independent off-shell string theory,''
Phys.\ Rev.\ D {\bf 47}, 3405 (1993)
[arXiv:hep-th/9210065].
}

\lref\WittenQY{
E.~Witten,
``On background independent open string field theory,''
Phys.\ Rev.\ D {\bf 46}, 5467 (1992)
[arXiv:hep-th/9208027].
}

\lref\GerasimovZP{
A.~A.~Gerasimov and S.~L.~Shatashvili,
``On exact tachyon potential in open string field theory,''
JHEP {\bf 0010}, 034 (2000)
[arXiv:hep-th/0009103].
}

\lref\SenMG{
A.~Sen,
``Non-BPS states and branes in string theory,''
arXiv:hep-th/9904207.
}

\lref\LambertHK{
N.~D.~Lambert and I.~Sachs,
``Tachyon dynamics and the effective action approximation,''
Phys.\ Rev.\ D {\bf 67}, 026005 (2003)
[arXiv:hep-th/0208217].
}

\lref\CallanMW{
C.~G.~Callan and I.~R.~Klebanov,
``Exact C = 1 boundary conformal field theories,''
Phys.\ Rev.\ Lett.\  {\bf 72}, 1968 (1994)
[arXiv:hep-th/9311092].
}

\lref\CallanUB{
C.~G.~Callan, I.~R.~Klebanov, A.~W.~Ludwig and J.~M.~Maldacena,
``Exact solution of a boundary conformal field theory,''
Nucl.\ Phys.\ B {\bf 422}, 417 (1994)
[arXiv:hep-th/9402113].
}

\lref\PolchinskiMY{
J.~Polchinski and L.~Thorlacius,
``Free Fermion Representation Of A Boundary Conformal Field Theory,''
Phys.\ Rev.\ D {\bf 50}, 622 (1994)
[arXiv:hep-th/9404008].
}

\lref\TseytlinMT{
A.~A.~Tseytlin,
``Sigma model approach to string theory effective actions with tachyons,''
J.\ Math.\ Phys.\  {\bf 42}, 2854 (2001)
[arXiv:hep-th/0011033].
}

\lref\TseytlinRR{
A.~A.~Tseytlin,
``Sigma Model Approach To String Theory,''
Int.\ J.\ Mod.\ Phys.\ A {\bf 4}, 1257 (1989).
}

\lref\FelderSV{
G.~N.~Felder, L.~Kofman and A.~Starobinsky,
``Caustics in tachyon matter and other Born-Infeld scalars,''
JHEP {\bf 0209}, 026 (2002)
[arXiv:hep-th/0208019].
}

\lref\HarveyWM{
J.~A.~Harvey, D.~Kutasov, E.~J.~Martinec and G.~Moore,
``Localized tachyons and RG flows,''
arXiv:hep-th/0111154.
}

\lref\MartinecTZ{
E.~J.~Martinec,
``Defects, decay, and dissipated states,''
arXiv:hep-th/0210231.
}

\lref\AdamsSV{
A.~Adams, J.~Polchinski and E.~Silverstein,
``Don't panic! Closed string tachyons in ALE space-times,''
JHEP {\bf 0110}, 029 (2001)
[arXiv:hep-th/0108075].
}

\lref\DavidVM{
J.~R.~David, M.~Gutperle, M.~Headrick and S.~Minwalla,
``Closed string tachyon condensation on twisted circles,''
JHEP {\bf 0202}, 041 (2002)
[arXiv:hep-th/0111212].
}

\lref\BergshoeffDQ{
E.~A.~Bergshoeff, M.~de Roo, T.~C.~de Wit, E.~Eyras and S.~Panda,
``T-duality and actions for non-BPS D-branes,''
JHEP {\bf 0005}, 009 (2000)
[arXiv:hep-th/0003221].
}

\lref\KlusonIY{
J.~Kluson,
``Proposal for non-BPS D-brane action,''
Phys.\ Rev.\ D {\bf 62}, 126003 (2000)
[arXiv:hep-th/0004106].
}

\lref\GarousiPV{
M.~R.~Garousi,
``Off-shell extension of S-matrix elements and tachyonic effective actions,''
arXiv:hep-th/0303239.
}

\lref\SchomerusUG{
V.~Schomerus,
``D-branes and deformation quantization,''
JHEP {\bf 9906}, 030 (1999)
[arXiv:hep-th/9903205].
}

\lref\SeibergVS{
N.~Seiberg and E.~Witten,
``String theory and noncommutative geometry,''
JHEP {\bf 9909}, 032 (1999)
[arXiv:hep-th/9908142].
}

\lref\MinahanTG{
J.~A.~Minahan and B.~Zwiebach,
``Gauge fields and fermions in tachyon effective field theories,''
JHEP {\bf 0102}, 034 (2001)
[arXiv:hep-th/0011226].
}

\lref\LambertFA{
N.~D.~Lambert and I.~Sachs,
``On higher derivative terms in tachyon effective actions,''
JHEP {\bf 0106}, 060 (2001)
[arXiv:hep-th/0104218].
}

\lref\TseytlinDJ{
A.~A.~Tseytlin,
``Born-Infeld action, supersymmetry and string theory,''
arXiv:hep-th/9908105.
}

\lref\GutperleXF{
M.~Gutperle and A.~Strominger,
``Timelike boundary Liouville theory,''
arXiv:hep-th/0301038.
}

\lref\GibbonsTV{
G.~Gibbons, K.~Hashimoto and P.~Yi,
``Tachyon condensates, Carrollian contraction of Lorentz group, and  fundamental strings,''
JHEP {\bf 0209}, 061 (2002)
[arXiv:hep-th/0209034].
}

\lref\SugimotoFP{
S.~Sugimoto and S.~Terashima,
``Tachyon matter in boundary string field theory,''
JHEP {\bf 0207}, 025 (2002)
[arXiv:hep-th/0205085].
}

\lref\MinahanIF{
J.~A.~Minahan,
``Rolling the tachyon in super BSFT,''
JHEP {\bf 0207}, 030 (2002)
[arXiv:hep-th/0205098].
}

\lref\IshidaFR{
A.~Ishida and S.~Uehara,
``Gauge fields on tachyon matter,''
Phys.\ Lett.\ B {\bf 544}, 353 (2002)
[arXiv:hep-th/0206102].
}

\lref\OhtaAC{
K.~Ohta and T.~Yokono,
``Gravitational approach to tachyon matter,''
Phys.\ Rev.\ D {\bf 66}, 125009 (2002)
[arXiv:hep-th/0207004].
}

\lref\GibbonsHF{
G.~W.~Gibbons, K.~Hori and P.~Yi,
``String fluid from unstable D-branes,''
Nucl.\ Phys.\ B {\bf 596}, 136 (2001)
[arXiv:hep-th/0009061].
}

\lref\KimHE{
C.~j.~Kim, H.~B.~Kim, Y.~b.~Kim and O.~K.~Kwon,
``Electromagnetic string fluid in rolling tachyon,''
JHEP {\bf 0303}, 008 (2003)
[arXiv:hep-th/0301076].
}

\lref\KimQZ{
C.~j.~Kim, H.~B.~Kim, Y.~b.~Kim and O.~K.~Kwon,
``Cosmology of rolling tachyon,''
arXiv:hep-th/0301142.
}

\rightline{EFI-03-13}
\Title{
\rightline{hep-th/0304045}}
{\vbox{\centerline{Tachyon Effective Actions}
\vskip 10pt \centerline{In Open String Theory}}}
\bigskip
\centerline{David Kutasov and Vasilis Niarchos}
\bigskip
\centerline{{\it Enrico Fermi Inst. and Dept. of Physics,
University of Chicago}}
\centerline{\it 5640 S. Ellis Ave., Chicago, IL 60637-1433, USA}
\bigskip\bigskip\bigskip
\noindent
We argue that the Dirac-Born-Infeld (DBI) action coupled to a tachyon, that
is known to reproduce some aspects of open string dynamics,
can be obtained from open string theory in a certain limit, which generalizes
the limit leading to the usual DBI action. This helps clarify which aspects of
the full open string theory are captured by this action.
\vfill

\Date{4/03}


\newsec{Introduction}

Recent work on the dynamics of unstable D-branes in string theory
has led to an effective action for the open string tachyon
$T$ and massless open string modes, $A_\mu$ (the gauge field
on the D-brane) and $Y^I$ (the scalar fields parametrizing the
location of the D-brane in the transverse directions)
\refs{\SenMD\GarousiTR\BergshoeffDQ\KlusonIY\GibbonsHF
\LambertFA\SenNU\SenIN\SenAN\SugimotoFP\MinahanIF\IshidaFR
\OhtaAC\LambertHK\GibbonsTV\KimHE\KimQZ-\GarousiPV}.
This action has the form\foot{We use the conventions $\alpha'=1$,
$\eta_{\mu\nu}=(-1,+1,\cdots,+1)$.}
\eqn\aaa{\eqalign{
\SS=&\int d^{p+1}\CL ~, \cr
\CL=&-V(T)\sqrt{-\det G} ~, \cr
}}
with $V(T)$ the tachyon potential (see below),
and
\eqn\bbb{G_{\mu\nu}=\eta_{\mu\nu}+\partial_\mu T\partial_\nu T
+\partial_\mu Y^I\partial_\nu Y^I+F_{\mu\nu} ~.}
The action \aaa\ is known to reproduce several non-trivial aspects
of open string dynamics, such as the following:
\item{(1)} Choosing \refs{\LeblondDB,\LambertZR}
\eqn\ccc{V(T)={1\over \cosh{\alpha T\over2}} ~,}
with $\alpha=1$ for the bosonic string, and $\alpha=\sqrt2$ for the non-BPS
D-brane in the superstring, one finds from \aaa\ the correct stress-tensor
$T_{\mu\nu}$ in homogenous tachyon condensation (the rolling tachyon solution
which starts at the top of the potential at $x^0\to-\infty$)
\refs{\SenNU,\LarsenWC,\LambertZR}.
\item{(2)} With the potential \ccc, one finds that the theory contains static solitonic
solutions corresponding to lower dimensional D-branes, with the correct tension.
\item{(3)} For the case of an unstable D-brane in type II string theory, one can construct
a codimension one BPS D-brane as a solitonic solution of \aaa. Small excitations of the
soliton correspond to massless fields, similar to $Y^I$ and $A_\mu$ in \bbb, and by
using \aaa\ one finds \SenTM\ that the effective action for these excitations of
the soliton is
precisely the DBI action\foot{Given by \aaa, \bbb\ with the tachyon $T$ set to zero;
see \TseytlinDJ\ for a review.}.
\item{(4)} Inhomogenous solutions of the equations of motion which follow from the
action \aaa\ encode non-trivial information about the decay of higher dimensional
branes into lower dimensional ones; in particular, they contain information about the
relative velocities of the lower dimensional branes created in the process of tachyon
condensation \refs{\FelderSV,\BerkoozJE}.
\item{(5)} For non-BPS D-branes in type II string theory, the potential
\ccc\ leads to the correct value of the mass of the tachyon on the D-brane (for the bosonic
string, this is not the case) \LambertZR.

\noindent
These and other successes lead one to believe that the action \aaa\ captures some class
of phenomena in the full classical open string theory. This action should presumably
be thought of as a generalization of the DBI action describing the gauge field $A_\mu$
and scalars $Y^I$ on the brane. The DBI action is valid in the full open string theory,
in situations where $F_{\mu\nu}$  and $\partial_\mu Y^I$ are arbitrary (\ie\ not necessarily
small), but slowly varying \TseytlinDJ. The question we would like to address in this
note is whether there exists a similar regime, in which the action
\aaa\ describes the interactions of the tachyon in the full open string theory.
We will argue that the answer is affirmative, and identify such a regime.

\newsec{An effective action for tachyons}

At first sight it seems difficult to incorporate the tachyon in an effective action such as the
DBI action, since its mass is of order the string scale. Solutions of the equations of motion,
$T(x^\mu)$, vary rapidly in spacetime, and in general one cannot decouple the tachyon from
other (non-tachyonic) modes with string scale masses.

To proceed, one can use the following fact. Consider a  homogenous tachyon
$T(x^0)$ in the open bosonic string\foot{We will discuss the generalization to
non-BPS branes in the superstring later.}. The general solution of the linearized
equation of motion for the tachyon is
\eqn\txo{T(x^0)=T_+e^{x^0}+T_-e^{-x^0} ~.}
It is known that \txo\ is an {\it exact} solution of the full open string equations
of motion\foot{To be precise, this is known to be the case in the Euclidean theory
obtained by taking $x^0\to ix$
\refs{\CallanMW,\CallanUB,\PolchinskiMY}, and is believed to be
the case in the Minkowski theory
as well.}. Thus, on-shell homogenous tachyons do not in fact couple to higher mass open
string modes. It is natural to expand around the exact solution \txo\ and study
tachyon profiles of the form
\eqn\tslow{T(x^\mu)=T_+(x^\mu) e^{x^0}+T_-(x^\mu)e^{-x^0} ~,}
where $T_\pm(x^\mu)$ are slowly varying on the string scale. What is the effective
action describing the dynamics of such slowly varying perturbations? This action should
have the property that arbitrary constant values of $T_+$ and $T_-$ correspond to a solution
of the equations of motion. It should describe the leading interactions in an expansion
in derivatives of $T_\pm (x^\mu)$. Such an action would be non-perturbative in $T$,
$\partial_\mu T$, since it would be valid for generic $T$ of the form \txo.
We will argue below that the action in question coincides (after a certain field
redefinition) with \aaa\ - \ccc.

Actually, if both $T_+$ and $T_-$ in \txo\ are non-vanishing, it is not obvious
that an effective action of the sort we want exists. The reason is that in this case,
the system is very far from the perturbative open string vacuum both at very early
$(x^0\to-\infty)$, and very late $(x^0\to\infty)$ time. Since the natural observables
in string theory are S-matrix elements of perturbative string modes, and the background
one gets as $T\to\pm\infty$ is not believed to contain any physical open string excitations,
it is not obvious that in this case one can make sense of the S-matrix, and therefore of
the action. In the case that either $T_+$ or $T_-$ vanishes, the situation is better.
Consider, say, the case\foot{This case was studied in \GutperleXF.}
\eqn\tplusx{T(x^0)=T_+e^{x^0}~.}
At early times, the tachyon goes to zero and the system approaches the perturbative
open string vacuum. Thus, one can define and study observables analogous to an S-matrix
as follows. The tachyon vertex operator in the bosonic string is
\eqn\tkw{T_{\vec k}=e^{i\vec k\cdot\vec x-w x^0};\;\;\vec k^2+w^2=1 ~.}
For $|\vec k| \ll 1$ one finds two solutions,
\eqn\wpm{w_\pm=\pm(1-\half\vec k^2)+O(\vec k^4)}
and thus the vertex operator \tkw\ takes the form
\eqn\tkpm{T_{\vec k}^{(\pm)}=e^{i\vec k\cdot\vec x-w_\pm  x^0} ~.}
Now, consider the correlation functions\foot{For now we restrict to amplitudes involving
only tachyons. We will comment on including massless fields below.}
\eqn\twonpt{\langle T_{\vec k_1}^{(+)}\cdots T_{\vec k_n}^{(+)}
T_{\vec p_1}^{(-)}\cdots T_{\vec p_m}^{(-)}\rangle_{T_+} ~.}
The subscript $T_+$ means that we are computing these correlation functions
in a background with a non-zero tachyon condensate \tplusx.
The correlation functions \twonpt\ vanish when
$\vec k_1=\cdots=\vec k_n=\vec p_1=\cdots=\vec p_m=0$. From the
spacetime point of view, this is due to the fact that \txo\ is an exact
solution of the full classical open string equations of motion. On the
worldsheet, the vanishing of these amplitudes is directly related to the
fact that the boundary perturbation $\lambda\int d\tau\cos x(\tau)$ is
truly marginal in the Euclidean case $(x^0\to ix)$.

The effective action we are after is the action that reproduces the correlation
functions \twonpt\ to leading order in $\vec k_i$, $\vec p_j$. In the next section
we will compute this effective action. We conclude this section with a few comments.

\item{(a)} To calculate \twonpt\ perturbatively in $T_+$, it is convenient to
continue to Euclidean space,
$x^0\to ix$. Then, the correlation functions \twonpt\ involve momentum
modes, whose momentum vectors are almost aligned with a particular, arbitrarily
chosen, axis in space.
\item{(b)} It might seem that the action we have introduced is not Poincare invariant,
since it involves a choice of preferred
direction in spacetime. This is not the case, since as $x^0\to-\infty$ the background
approaches a Poincare invariant one (the open string vacuum), and the action should
be valid for arbitrary perturbations away from this vacuum obtained from \tplusx\
by a Poincare transformation. In other words, the apparent breaking of Poincare symmetry
in \tplusx\ is spontaneous.
\item{(c)} One can think of the effective action we have introduced as
a special case of a more general construction, which includes the usual DBI action as
another special case. In the $\sigma$-model approach to string theory
(see, for example, \refs{\TseytlinRR,\TseytlinMT}), one thinks
of the configuration space of the theory as the space of (in general non-conformal)
worldsheet theories, with conformal theories corresponding to solutions of the
spacetime equations of motion. The coupling $T_+$ in \tplusx\ parametrizes a line of fixed
points of the worldsheet renormalization group (RG), or classical solutions of
the spacetime action\foot{Solutions with different $T_+\not=0$ are related by time
translation and are thus equivalent.}.
The effective action we have introduced describes infinitesimal deviations from this
line of fixed points. Since turning on a large constant $T_+$ in \tslow\ clearly does not
take us away from this line, the size of the deviation from the line of fixed points is
governed by the size of the {\it derivatives} of $T_+(x^\mu)$, and by $T_-$.
This is analogous to the
situation in the DBI case, where there is a surface of fixed points of the worldsheet RG
corresponding to constant $F_{\mu\nu}$, $\partial_\mu Y^I$. The DBI action governs
small fluctuations away from this surface of fixed points.
\item{(d)} It is easy to generalize the considerations of this section to the case of non-BPS
branes in the superstring. The homogenous tachyon takes in this case the
form
\eqn\txosup{T(x^0)=T_+\exp({1\over\sqrt2}x^0)+T_-\exp(-{1\over\sqrt2}x^0) ~.}
This is also the vertex operator in the $-1$ picture. Eq. \txosup\ corresponds
again to an exactly marginal perturbation of the worldsheet theory, and thus
to an exact solution of the spacetime equations of motion. One can study small
fluctuations around the solution with $T_+\not=0$, $T_-=0$, as in \tslow,
and define an effective action
for these fluctuations. This effective action should reproduce the leading
small $\vec k$ behavior in correlation functions of the analogs of the
vertex operators \tkpm. In the fermionic case one has $(-1)$-picture
vertex operators
\eqn\tkminusone{T_{\vec k}^{(\pm)}=e^{i\vec k\cdot\vec x-w_\pm  x^0} ~,}
with
\eqn\omsuppm{w_\pm=\pm{1\over\sqrt{2}}(1-\vec k^2)+O(\vec k^4) ~.}
The corresponding $0$-picture vertex operators are
\eqn\tkzero{T_{\vec k}^{(\pm)}=i(\vec k\cdot\vec\psi-w_\pm\psi^0)e^{i\vec k\cdot\vec x-w_\pm  x^0}}
and the correlation function \twonpt\ contains $2n-2$ $0$-picture vertex operators and two
$(-1)$-picture ones.
\item{(e)} There are some well known ambiguities associated with going
from on-shell S-matrix elements to off-shell actions. One is the freedom
to perform field redefinitions $T\to f(T,\partial_\mu T,\partial_\mu
\partial_\nu T,\cdots)$. Another is the freedom to use the equations
of motion at lower order in $T$ to change higher order terms in the
Lagrangian. For example, for a field $T$ of unit mass, a cubic vertex
$g_3 T^3$ gives the same on-shell three point function as the derivative
interaction $g_3(\partial_\mu\partial^\mu T) T^2$. We will arrive
at a specific form of the action by fixing all these ambiguities in a particular
way, but of course one can write the action in a different form by
using them.

\newsec{Computing the effective action}

In this section we will compute the effective Lagrangian for the tachyon
discussed above. We will require the Lagrangian to be symmetric under $T\to -T$.
In the fermionic case (the non-BPS brane), this is due to the standard
$Z_2$ symmetry of the theory, $\psi^\mu\to-\psi^\mu$,
under which the open string tachyon is odd.
In the bosonic case, one should in principle start from a Lagrangian without
such a symmetry but, as we mention below, imposing one of the conditions
that $\CL$ should satisfy leads to a Lagrangian even under $T\to -T$. Thus, we
impose this symmetry from the outset in the bosonic case as well.

Since the action is designed to reproduce only the leading terms in the S-matrix
elements \twonpt\ as $\vec k_i, \vec p_j\to 0$, we can furthermore take the Lagrangian
$\CL$ to depend on $T$ and $\partial_\mu T$ only,
\eqn\ltmut{\CL=\CL(T,\partial_\mu T) ~.}
This is essentially the statement that Lagrangians of the form \ltmut\
have a sufficient number of free parameters to match the leading terms
in \twonpt\ for all $n,m$. We will return to this point and will make it
more precise in section 4. For now, we will take the fact that we can bring
$\CL$ to the form \ltmut\ for granted, and proceed.

Note that \ltmut\ partially fixes the field redefinition ambiguity
mentioned at the end of section 2, but it leaves a residual freedom of taking
\eqn\tfttwo{T\to Tf(T^2) ~,}
where $f(x)=1+C_1x+C_2x^2+\cdots$.

To summarize, one expects the Lagrangian to take the form

\eqn\lexp{\CL=\sum_{n=0}^\infty\CL_{2n}(T,\partial_\mu T) ~,}
where $\CL_{2n}$ includes all the terms that go like $T^{2n}$,
\eqn\ltwon{\CL_{2n}=\sum_{l=0}^na_l^{(n)}(\partial_\mu T\partial^\mu T)^l
T^{2(n-l)} ~.}
It is important to emphasize that in the preceeding discussion
we have assumed that the effective Lagrangian \ltmut\ is analytic
around $T=0$. This is in fact not guaranteed, and we will see that this
assumption fails in some cases.

Under the assumptions outlined above, the problem of determining
the Lagrangian reduces to computing the constants $a_l^{(n)}$.
A non-trivial constraint is that the equations of motion that
follow from the Lagrangian \lexp, \ltwon\ should allow\foot{See
\refs{\LambertFA,\LambertHK} for a related discussion.} the solution
\txo. Since \txo\ should be a solution for arbitrary (constant)
$T_\pm$, the equations of motion that follow from \ltwon\ should
allow this solution for each $n$ separately.

Varying $\CL_{2n}$, one finds the equation of motion
\eqn\eqofmot{\sum_{l=1}^nla_l^{(n)}\partial^\mu\left[
(\partial_\lambda T\partial^\lambda T)^{l-1}(\partial_\mu T)
T^{2(n-l)}\right]=
\sum_{l=0}^n(n-l)a_l^{(n)}T^{2(n-l)-1}(\partial_\mu T\partial^\mu T)^l ~.}
Plugging \txo\ in \eqofmot\ leads to a recursion relation for the
$a_l^{(n)}$,
\eqn\recal{a_{l+1}^{(n)}={(n-l)(2l-1)\over (2l+1)(l+1)}a_l^{(n)} ~,}
with the solution
\eqn\alone{a_l^{(n)}={(n-1)! \over (n-l)!l!(2l-1)} a_1^{(n)} ~.}
We see that all couplings in \ltwon\ are fixed in terms of any one
of them by the requirement that \txo\ be a solution of the equations
of motion \eqofmot. It should be mentioned that if one starts with a
Lagrangian without the symmetry $T\to -T$, the requirement that \txo\
is a solution implies that all terms odd under this symmetry must vanish.

To fix $\CL$, we have to compute the remaining unknown coefficients,
$a_1^{(n)}$. Note that the requirement that \txo\ be a solution fixes
the field redefinition freedom \tfttwo, and we expect to find a unique
solution for the Lagrangian \lexp.

In order to compute the couplings $a_1^{(n)}$ \alone, one can proceed
as follows. On general grounds, one expects that the on-shell spacetime
action should be equal to the disk partition sum
\refs{\TseytlinRR,\WittenQY,\TseytlinMT}. Both the spacetime
action and the disk partition sum involve an integral over $x^0$, and
in both of them there is a natural object that can be defined by stripping
off the integral over $x^0$. In the case of the spacetime action,
the resulting object is the on-shell Lagrangian \lexp. In the case of
the worldsheet partition sum, it is the disk path integral over the
non-zero modes of $x^0$, with the zero mode unintegrated,
$Z'(x^0)$. It is natural to conjecture that these two objects
are equal to each other,
\eqn\lzprime{\CL_{\rm on-shell}(x^0)=Z'(x^0) ~.}
This assumption was used successfully in \LarsenWC, and we will
use it here as well. Using \lzprime, one can fix all the
$a_1^{(n)}$ as follows. Substitute the solution \tplusx\
into the Lagrangian \lexp, \ltwon, and compare the resulting
function of $x^0$ to that obtained by evaluating $Z'(x^0)$ \LarsenWC
\eqn\zpeval{Z'(x^0)={1\over 1+T_+ e^{x^0}} ~.}
In doing that, one encounters a surprise. The on-shell
Lagrangian \lexp, \ltwon, \tplusx\ only involves terms
that go like $\exp(2nx^0)$, $n=0,1,2,3,\cdots$, while $Z'$
\zpeval\ also has in its expansion odd powers of $\exp(x^0)$.
Thus, it seems that one of the assumptions that went into the
analysis above must be incorrect. We will soon see that the
problematic assumption is that of analyticity of
$\CL(T,\partial_\mu T)$ near $T=0$, but for now let us set
this problem aside and turn to the fermionic string (the non-BPS
brane in type II), which as we will see is easier to understand.

First note that the analysis leading to \alone\ is slightly modified
in this case, since the solution we want is not \txo\ but \txosup.
It is easy to see that the correct form of \alone\ for the fermionic
case is
\eqn\aloneferm{a_l^{(n)}={(n-1)!2^{l-1} \over (n-l)!l!(2l-1)} a_1^{(n)} ~.}
We can now attempt to fix the coefficients $a_1^{(n)}$
by computing the effective Lagrangian
\lexp, \ltwon\ for the solution $T(x^0)=T_+\exp(x^0/\sqrt2)$ (see \txosup),
and comparing it to the disk partition sum $Z'(x^0)$. The latter was computed
in \LarsenWC:
\eqn\zprimesup{Z'={1\over 1+\frac{1}{2} T_+^2 e^{\sqrt{2}x^0}} ~.}
We see that in this case the expansion of $Z'$ involves only
even powers of $T(x^0)$, and we can use \lzprime\ to determine
$a_1^{(n)}$. One finds the following result:
\eqn\alnans{a_l^{(n)}=-{(-1)^n2^{l-2n}(2n-1)!!\over l! (n-l)! (2l-1)} ~,}
where we have used the identity
\eqn\idone{\sum_{s=0}^n \frac{(-1)^s}{s!(n-s)!(2s-1)}=-\frac{2^n}{(2n-1)!!} ~.}
As a check, \alnans\ can be easily verified to satisfy \aloneferm.
Plugging \alnans\ into \lexp, \ltwon\ one finds\foot{With the help
of another identity, $$\sum_{s=0}^{n-l} 2^{-s} \frac{(2(l+s)-3)!!}{s!}=
\frac{2^{l-n}(2n-1)!!}{(n-l)!(2l-1)} ~.$$}
\eqn\ferlag{\CL=-{1\over 1+\half T^2}
\sqrt{1+\half T^2+\partial_\mu T\partial^\mu T}~.}
Finally, making the redefinition
\eqn\tredef{{T\over\sqrt2}=\sinh{\tilde T\over\sqrt2} ~,}
which is a transformation of the form \tfttwo, one finds
\eqn\lsupfinal{
\CL=-{1\over\cosh{\tilde T\over\sqrt2}}\sqrt{1+\partial_\mu \tilde
T\partial^\mu \tilde T} ~.}
This is exactly the tachyonic part of the Lagrangian \aaa\ - \ccc.

Having understood the string theory origin of the
(tachyon part of the) Lagrangian \aaa\ for the non-BPS D-brane
in type II string theory, we return to the bosonic case.
Let us assume that the Lagrangian \aaa, \bbb\ with the potential
\ccc\ is still correct in this case, \ie\ that in some definition
of the tachyon field, $\tilde T$, the Lagrangian is
\eqn\ltildebos{
\CL=-{1\over\cosh{\tilde T\over2}}\sqrt{1+\partial_\mu \tilde
T\partial^\mu \tilde T} ~.}
The equation of motion of \ltildebos\ has a solution
\eqn\halfb{\sinh {\tilde T\over2}=\exp(\half x^0)}
whose energy density is the same as that of the original D-brane. It is
easy to see that this solution corresponds in the parametrization
\txo\ to \tplusx. Thus, the parametrizations \halfb, \tplusx\
are related by a map of the form
\eqn\tttilde{T=C\sinh^2{\tilde T\over 2} ~,}
with $C$ some constant. In particular, the map is non-analytic
near $T=0$. One has:
\eqn\ttexp{T=a\tilde T^2+b\tilde T^4+\cdots ~.}
The Lagrangian \ltildebos, which is analytic in $\tilde T$, corresponds
in terms of $T$ to a non-analytic Lagrangian
\eqn\lnonanal{
\CL={\rm const}-{1\over2}(\partial_\mu\tilde T)^2+{1\over 8}
\tilde T^2+\cdots={\rm const}-{1\over 8a}{(\partial_\mu T)^2\over T}
+{1\over 8a}T+\cdots ~.}
We see that, as suggested by the expansion of the disk partition sum \zpeval,
the on-shell Lagrangian does contain odd powers of $T\simeq \exp(x^0)$,
and these are due to the non-analytic structure of the effective Lagrangian
for $T$ near $T=0$. One can in fact check that plugging in the solution
\halfb\ into the Lagrangian \ltildebos\ leads to a result which agrees with
the disk partition sum \zpeval, as one would expect from \lzprime.

\newsec{Discussion}

In the previous sections we have argued that the action \aaa\ - \ccc\
arises from string theory in a particular limit, in which one
considers slowly varying $T_\pm(x^\mu)$ in \tslow, \txosup\
(in the bosonic and fermionic cases, respectively), expanded around a solution
with $T_+\not=0$, $T_-=0$. In this section
we will discuss in more detail the regime of validity of the action
\aaa\ - \ccc, comment on possible extensions of our results, and discuss
some consequences of our analysis for the issues mentioned in the introduction.

What is the precise string theory question, the answer to which is the action
\aaa\ - \ccc? Consider, for example, the scattering amplitude of tachyons on
a non-BPS brane, \tkminusone\ - \tkzero,
\eqn\twonpptt{\langle T^{(+)}_{\vec k_1}\cdots T^{(+)}_{\vec k_n}
T^{(-)}_{\vec p_1}\cdots T^{(-)}_{\vec p_n}\rangle}
in the limit $|\vec k_i|, |\vec p_j| \ll 1$. At $\vec k_i=\vec p_j=0$,
this amplitude vanishes, since \txosup\ is an exact solution
of the full classical open string equations of motion. Thus, the leading
non-vanishing terms in \twonpptt\ scale like momentum (various combinations
of $\vec k_i$, $\vec p_j$) squared. What we have shown in sections 2 and 3
is that \aaa\ - \ccc\ is the action that reproduces these momentum squared
terms in the $2n$-point function \twonpptt. This way of thinking about it
also explains why one can choose the Lagrangian to depend only on $T$ and
$\partial_\mu T$, as we have in \ltmut.

Indeed, as mentioned in section 3, Lagrangians
of the form \ltmut\ have enough free parameters to incorporate the fact that \twonpptt\
vanishes when all $\vec k_i=\vec p_j=0$, and to parametrize the most general
possible $O((\vec k, \vec p)^2)$ terms in \twonpptt.
The vanishing of the amplitude at zero momentum
corresponds to the requirement that \txosup\ is an exact solution
of the equations of motion, and leads to the determination of all terms at order
$T^{2n}$ in terms of one constant (see \aloneferm). The fact that the remaining
single free parameter at each order, $a_1^{(n)}$, is sufficient to parametrize
the most general momentum squared terms in \twonpptt\
can be seen in the following way.

The $2n$-point amplitude \twonpptt\ has
two different types of contributions: vertices with less than $2n$ external legs connected
by propagators, and one-particle-irreducible vertices with $2n$ external legs.
To determine the form of the effective action, we are only interested in the 1PI contributions. These are local, i.e.\ polynomial,
in the momenta $\vec k_i$ and $\vec p_j$ and their form is severely constrained
by symmetries, as follows. The most general local quadratic polynomial in $\vec k_i$, $\vec p_j$,
compatible with rotation invariance can be written in the form
\eqn\quadratic{\sum_{i,j} a_{ij}\vec k_i\cdot\vec k_j+\sum_{i,j} b_{ij}\vec p_i\cdot\vec p_j
+\sum_{i,j} c_{ij}\vec k_i\cdot\vec p_j ~,}
where $a_{ij}$, $b_{ij}$, $c_{ij}$ are free parameters that have to satisfy additional symmetry constraints.
First, since the amplitude \twonpptt\ is by definition symmetric
under interchange of any of the $T^{(+)}$'s and separately
under interchange of any of the $T^{(-)}$'s, (i.e.\ it is
completely symmetric under interchange of the $\{\vec k_i\}$,
and separately symmetric under interchange of the $\{\vec p_j\}$) it can be written as
\eqn\abc{a_1\sum_i |\vec k_i|^2+a_2\sum_{i\neq j} \vec k_i\cdot\vec k_j+
b_1\sum_i |\vec p_i|^2+b_2\sum_{i\neq j} \vec p_i\cdot\vec p_j+
c\sum_{i,j} \vec k_i\cdot\vec p_j ~.}
Since $\sum_i \vec k_i+\sum_j\vec p_j=0$ by momentum conservation,
the last term (proportional to $c$) is not independent of the other terms,
and we can set $c=0$.
Furthermore, the symmetry $(\vec x, t)\rightarrow(-\vec x, -t)$ implies that
the amplitude should also be invariant under interchange of the $T^{(+)}$'s
with the $T^{(-)}$'s, i.e.\ it should be invariant under the interchange of
$\{\vec k_i\}$ with $\{\vec p_j\}$. This implies that $a_1=b_1$ and $a_2=b_2$ and
we are left with only two independent constants, $a_1$ and $a_2$.
The fact that \twonpptt\ is an amplitude in a Lorentz invariant theory, and thus
we can write it in terms of Mandelstam invariants, implies that
$a_1$ and $a_2$ are also related. Altogether, we conclude
that to quadratic order in momenta the 1PI part of
the $2n$-point function \twonpptt\ is unique
up to one free coefficient. The action that was constructed in sections 2 and 3
also has one free coefficient at each order,
and the argument leading to \alnans\ determines it.

Since the action \aaa\ - \ccc\ describes the leading terms in the $2n$-point function
\twonpptt, one can also use it to describe scattering amplitudes in the presence of a
non-zero $T_+$, as in \twonpt, perturbatively in $T_+$.
Note that one cannot use it when both $T_+$ and $T_-$ in \txosup\ are non-zero, since then
the spectrum of scaling dimensions changes (it is no longer true that
$\Delta(e^{wx_0})=w^2$). This is the worldsheet manifestation of the fact
(mentioned in section 2) that in this case the system is far from the open string
vacuum both at early and at late times.

We now move on to possible extensions of our work.
One natural extension is to add the gauge field on the D-brane
$A_\mu$, and scalars $Y^I$, to derive the full action \aaa\ - \ccc. This should
be possible along the lines of comment (c) at the end of section 2. To construct
the tachyon action \lsupfinal\ we utilized the existence of exact
classical solutions (or, equivalently, manifolds of fixed points of the
worldsheet boundary RG) labeled by $T_+$. We also noted that constant $F_{\mu\nu}$,
$\partial_\mu Y^I$, correspond to surfaces of solutions as well, and studying the
vicinity of these surfaces leads to the DBI action.

One can combine the two observations and study surfaces of solutions
labeled by constant $F_{\mu\nu}$, $\partial_\mu Y^I$, and $T_+$. This is
particularly simple when $F_{\mu0}=\partial_0 Y^I=0$, since then the solution
for the tachyon \txo, \txosup\ is not modified by the expectation values of
the massless fields. More generally, one has to replace \txo, \txosup\ by a solution
of the tachyon equations of motion in the open string metric, but one still expects
to have a surface of solutions labeled by constant expectation values of
$F_{\mu\nu}$, $\partial_\mu Y^I$, and $T_+$. The full action describing
slowly varying $F_{\mu\nu}$, $\partial_\mu Y^I$, $T_+$ (and small and slowly
varying $T_-$) is very likely given by \aaa\ - \ccc.  This leads to a uniform
treatment of the tachyon and massless fields: the action \aaa\ - \ccc\ describes
the physics in the vicinity of the surface of exact solutions corresponding to
constant $F_{\mu\nu}$, $\partial_\mu Y^I$, $T_+$.

Another generalization is to couple the action \aaa\ - \ccc\ to massless closed
strings. For (NS,NS) sector closed strings, this is expected to lead to a structure
similar to the usual DBI action. The tachyon couplings to massless (R,R) sector
fields are an interesting open problem.

In the introduction we mentioned that the action \aaa\ - \ccc\ seems to reproduce some aspects
of the dynamics of the full open string theory like properties of rolling tachyon and other
solutions. Our understanding of the role of this action in string theory should help clarify
which aspects of the full problem should be captured by this action, and which should not.
We next briefly comment on this issue in the context of the points mentioned in section 1.

The fact that the action \aaa\ - \ccc\ should reproduce the correct stress-tensor in homogenous
tachyon condensation should be clear from the point of view of our analysis. This
is the statement that the one point function of a zero momentum graviton in the full string theory
is the same as the expectation value of the stress-tensor corresponding to \aaa\ on the solution \txo,
\txosup. Clearly, this one point function probes an infinitesimal deviation from the solution that the
action \aaa\ is designed to describe, and thus the result obtained from \aaa\ must agree with that
obtained in the full string theory. Indeed, this is known to be the case for the solution \tplusx.

For the general solutions \txo, \txosup, with both $T_+$ and $T_-$ non-zero, it
is known that the stress tensor computed in the field theory \aaa\ does not precisely
reproduce that computed in the full string theory \SenQA. For example, in the case of
non-BPS branes in the superstring, when the energy density $E$ is smaller than the D-brane
tension, the tachyon effective action \ferlag\ gives \LambertZR\ the stress-energy
tensor\foot{We have set the D-brane tension to one.}
\eqn\fieldstress{T_{00}=E, ~ ~ T_{ij}=-\frac{1}{E}
\frac{1}{1+\frac{u^2}{2} \cosh^2 \big ( \frac{x^0}{\sqrt{2}} \big)}\delta_{ij}=
-\delta_{ij}
\Big(\frac{1}{1+ce^{\sqrt{2}x^0}}+\frac{1}{1+ce^{-\sqrt{2}x^0}}-1
\Big)
~,}
where $T_+=T_-=\frac{u}{2}$ in \txosup, $c=\frac{b-1}{b+1}$,
and $b^2=1+\frac{u^2}{2}={1\over E^2}$.
The exact open string calculation gives \SenIN\
\eqn\stringstress{T_{00}=\frac{1}{2}(1+\cos u), ~ ~ T_{ij}=-f(x^0)
\delta_{ij} ~,}
with
\eqn\fx{f(x^0)=\frac{1}{1+\half\sin^2(\frac{u}{2}) ~ e^{\sqrt{2}x^0}}+
\frac{1}{1+\half\sin^2(\frac{u}{2}) ~ e^{-\sqrt{2}x^0}}-1 ~.}
The two results \fieldstress\ and \stringstress\ have the same rough form, and agree
to leading order in an expansion in $u$, but the
detailed dependence on the strength of the tachyon field is not the same. This is
consistent with our discussion in section 2. Note that as pointed out in
\SenQA, even in this case the effective field theory \aaa\ - \ccc\ is still valid
at late times. This is very natural from the point of view of the discussion in section
2, since at late times one can replace the interaction \txo\ by \tplusx, with a renormalized
value of the coupling $T_+$ (this renormalization is the origin of the periodic dependence
on the tachyon in \fx, from this point of view).

We mentioned in section 1 that the action \aaa\ describes correctly lower dimensional D-branes,
which correspond to solitons of this action. This is easy to understand from the point of view
of our analysis. As explained in \HarveyNA, one way to describe codimension one
D-branes in the full string theory is to turn on a tachyon with the profile
\eqn\tachsol{T(x)=e^{wx^0}\cos kx ~.}
(in \HarveyNA, $\exp(wx^0)$ is replaced by a scale dependent coupling, but this
is an inessential difference). In the bosonic string, one finds at late times
codimension one D-branes located at the minima of \tachsol, while in the
fermionic case one finds codimension one BPS D-branes located at the zeroes of
\tachsol.

As mentioned in \HarveyNA, taking $k\to 0$ in \tachsol\ allows one to focus on a
single codimension
one D-brane, with the other branes sent to infinity in the limit. Perturbations of the form \tachsol\
with $k\to 0$ (which correspond to $T_+(x)\propto \cos kx$ in \tslow, \txosup) are precisely what
the action \aaa\ is designed to describe. $T_+(x)$ is slowly varying and the perturbation never
takes one far from the surface of exact solutions corresponding to constant $T_+$. Therefore,
one {\it expects} \aaa\ to give the correct tension of the codimension one brane, and (for the non-BPS
brane in type II) to give the Dirac-Born-Infeld action as the action describing small excitations of
the soliton\foot{In the bosonic case, one expects to get an action of the form \aaa\ on the soliton.}.
In the full string theory, the approximation that gives rise to \aaa\ on the original non-BPS brane
is exactly the same as the approximation that gives rise to the DBI action on the
BPS brane. One goes smoothly into the other under the tachyon condensation
\tachsol\ with $k \ll 1$.

What happens if we turn on a perturbation \tachsol\ with $k\simeq 1$? Since $T_+(x)$ is no
longer slowly varying, the action \aaa\ need not describe the full open string dynamics in this
case. However, as discussed in \HarveyNA, at late times one actually expects the
worldsheet
operator $\cos kx$ to approach the identity operator, and the full boundary CFT is thus
expected to approach the surface of solutions $T_+={\rm constant}$, on which the action \aaa\
is reliable. Thus, in situations like this, one expects the dynamics to be well approximated by
\aaa\ at late times, when the tachyon is large. This agrees with the discussion
in \SenQA.

General inhomogenous solutions of the equation of motion of the tachyon action \aaa\ in
$1+1$ dimensions were analyzed in \FelderSV. It was found that generic solutions
develop caustics at a finite time. It was then argued in \SenQA\ that the
effective action \aaa\ breaks down in these situations, since the tachyon
develops large gradients. In \BerkoozJE\ it was pointed out
that the presence/absence of caustics in the solutions is directly related to the dynamics of the
$D0$-branes that the $D1$-brane decomposes into via tachyon condensation. If these $D0$-branes
are at rest relative to each other, the late time solution is smooth. If they have non-zero relative velocities,
caustics appear and \aaa\ breaks down. From the point of view of our analysis here, this is very
reasonable. The situation with vanishing relative velocities can be Lorentz transformed\foot{This
is true for equidistant $D0$-branes at late times. Non-equidistant  $D0$-branes at rest are
described by a simple generalization of \tachsol.} to \tachsol,
which as we already explained is well described by \aaa. When the relative velocities are non-zero,
it is easy to see that the system is getting farther and farther away from the surface of solutions
$T_+={\rm constant}$ at late times. Thus, the action \aaa\ is not expected to be reliable in this case,
and the caustics presumably signal its breakdown.

Finally, we mentioned in section 1 that the action \aaa\ - \ccc\ reproduces the correct mass of the
tachyon on the non-BPS D-brane in type II, while it does not give the correct tachyon mass in the
bosonic string. From the point of view of our analysis, the fact that the mass is reproduced in the type
II case is true by construction, while the disagreement in the bosonic string is understood to be due to
the non-analytic relation \tttilde\ between the open string tachyon $T$ and the field that appears in \aaa,
which is really $\tilde T$. Taking the map \tttilde\ into account, one in fact finds the correct tachyon
mass (again, by construction).

The action \aaa\ - \ccc\ is quite different from the actions that were found in boundary string
field theory (BSFT) \refs{\GerasimovZP,\KutasovQP,\KutasovAQ}. It is sometimes said
that these actions might be related by field redefinitions, which involve
derivatives of $T$. From the point of view presented here,
it is clear why these actions look so different. The actions
computed in BSFT are valid far off the mass-shell of $T$, \eg\ for $T\simeq a+ux^2$ in the bosonic
string. The action \aaa\ - \ccc\ on the other hand is valid for approximately on-shell configurations,
such as \tkw, \wpm. The two actions have rather different regimes of validity, and it is not
surprising that they look different.

\lref\StromingerFN{
A.~Strominger and T.~Takayanagi,
``Correlators in timelike bulk Liouville theory,''
arXiv:hep-th/0303221.
}

\lref\VafaRA{
C.~Vafa,
``Mirror symmetry and closed string tachyon condensation,''
arXiv:hep-th/0111051.
}

The general point of view on effective actions presented here
might be useful for thinking about other related
problems as well. For example, one can ask whether it is possible to generalize the analysis to the
study of  closed string tachyons. In order to do that, one needs to identify an analog of the exact
solution \txo\ for this case. As a naive attempt, one may try to study the worldsheet theory in the presence
of a perturbation
\eqn\closedcos{\delta\CL_{\rm ws}=\lambda\cosh 2x^0 ~.}
Unfortunately, unlike its open string analog, this perturbation is not expected to be truly marginal. Analytically
continuing $x^0\to ix$, one finds the Sine-Gordon interaction, which is marginally relevant. The central
charge goes to zero in the IR (the worldsheet field $x$ becomes massive).
In the string theory context, this means that turning on $\lambda$ leads to a
large backreaction of the metric and dilaton at late times, and one needs to understand it before proceeding.
It is possible that, as was suggested recently in \StromingerFN, the perturbation
\eqn\lflflf{\delta\CL_{\rm ws}=\lambda\exp(2x^0)}
is exactly marginal, in which case one could use it
as a closed string analog of \tplusx, although it is not clear what suppresses the
backreaction of the metric and dilaton to the stress-tensor of the tachyon.
Another natural arena in which effective actions such as \aaa\ might be useful is localized closed
string tachyon condensation \refs{\AdamsSV\VafaRA\HarveyWM\DavidVM-\MartinecTZ}, where the
backreaction is milder, and in particular the central charge does not change.

\vskip 1cm
\noindent{\bf Acknowledgments:}
We are grateful to B. Craps, F. Larsen, E. Martinec and E. Rabinovici
for discussions. This work was supported in part by DOE grant
DE-FG02-90ER-40560.

\listrefs
\end